\documentclass{article}

\usepackage{microtype}
\usepackage{graphicx}
\usepackage{subfigure}
\usepackage{booktabs} 

\usepackage{hyperref}

\usepackage[accepted]{sty}

\usepackage{times}
\usepackage{helvet}
\usepackage{courier}
\usepackage{subfigure} 
\usepackage{cancel}
\RequirePackage{amsthm,amsmath}
\usepackage{graphicx, amssymb, mathrsfs,amsfonts,url, bm}

\newcommand{\real}{\mathbb{R}}

\newcommand{\gap}{\,\,\,\,\,\,\,\,}

\newcommand{\bff}{\mathbf{f}}
\newcommand{\bg}{\mathbf{g}}
\newcommand{\bh}{\mathbf{h}}
\newcommand{\bX}{\mathbf{X}}
\newcommand{\bx}{\mathbf{x}}
\newcommand{\by}{\mathbf{y}}

\newcommand{\bw}{\mathbf{w}}

\newcommand{\titlethis}{Reducing overestimating and underestimating volatility via the augmented blending-ARCH model}

\icmltitlerunning{\titlethis}

\begin{document}

\twocolumn[
\icmltitle{\titlethis}

\begin{icmlauthorlist}
\icmlauthor{Jun Lu}{te}
\icmlauthor{\gap\gap Shao Yi}{}
\end{icmlauthorlist}

\icmlaffiliation{te}{Correspondence to: Jun Lu $<$jun.lu.locky@gmail.com$>$. Copyright 2022 by the author(s)/owner(s). March 16, 2022}


\vskip 0.3in
]



\printAffiliationsAndNotice{}  

\begin{abstract}
SVR-GARCH model tends to ``backward eavesdrop" when forecasting the financial time series volatility in which case it tends to simply produce the prediction by deviating the previous volatility. 
Though the SVR-GARCH model has achieved good performance in terms of various performance measurements, trading opportunities, peak or trough behaviors in the time series are all hampered by underestimating or overestimating the volatility.
We propose a blending ARCH (BARCH) and an augmented BARCH (aBARCH) model to overcome this kind of problem and make the prediction towards better peak or trough behaviors. The method is illustrated using real data sets including SH300 and S\&P500. The empirical results obtained suggest that the augmented  and blending models improve the volatility forecasting ability.

\end{abstract}
\section{Introduction}
\noindent In quantitative finance, the assert returns are usually modeled by the normal distribution since its form is normal (hence the name normal distribution) \citep{levy2004asset,wirjanto2009applications, lu2022exploring}. Due to its normal form, the normal distribution cannot model the fat tails (leptokurtosis) and asymmetry (skewness) of the asset returns. \citet{choi2008asymmetric} suggested the SU-normal distribution to describe the two non-normal features embedded in financial time series. And recently, a polynomial adjusted Student-t distribution is proposed to model asset returns with heavy-tailed and skewed distributions \citep{leon2021polynomial}.

Volatility is a measure of the degree of fluctuation of financial return and is a proxy for risk which is incorporated into the famous Sharpe ratio and information ratio to measure the performance of a quantitative strategy on the asset \citep{sharpe1966mutual, sharpe1994sharpe, goodwin1998information}. And even the volatility is used to quantify the premium (option price) in both option call and option put; the higher the volatility, the higher the option price \citep{haug2007complete, natenberg2014option}. Conditional heteroscedastic models such as the autoregressive conditional heteroscedasticity (ARCH) \citep{engle1982autoregressive}, generalized ARCH (GARCH) \citep{bollerslev1986generalized}, exponential GARCH (EGARCH) \citep{nelson1991conditional}, and GJR \citep{glosten1993relation} can model the volatility time series that 
exhibit time-varying volatility and volatility clustering, i.e., periods of swings interspersed with periods of relative calm. However, empirical studies show that these models have low forecasting accuracy \citep{jorion1995predicting, choudhry2008forecasting, bezerra2017volatility}. 

In recent years, machine learning methods have been employed to provide superior results in many time series problems, e.g., flight ticket prediction \citep{lu2017machine, rajankar2019survey}, and metabolic pathway dynamics \citep{costello2018machine}.
Machine learning-based methods have also been proposed to improve the performance since it can capture nonlinear features hidden in the time series such as leptokurtosis, asymmetry, volatility clustering, and momentum, e.g., empirical test shows support vector regression (SVR) has superior results than GARCH models \citep{perez2003estimating, chen2010forecasting, li2014estimating, bezerra2017volatility}. 

Empirical evidence shows that there are oscillations between several regimes
in the financial market, in which the overall distribution of returns is a mixture of
two or more than two states of normal \citep{guidolin2011markov, levy2015portfolio}. 
\citet{bezerra2017volatility} proposed an SVR-GARCH model that captures the regime-switching behavior and performs better than existing methods since SVR is a kernel-based method and the Gaussian kernel can be decomposed into an infinite mixture of polynomials \citep{lu2021rigorous}. However, a pictorial analysis on the volatility series via the SVR-GARCH model shows that the model tends to ``backward eavesdrop", that is, the model simply reports the volatility by the previous value (in a sense of deviation). Moreover, a pictorial analysis on the predicted series shows that the SVR-GARCH model tends to underestimate the volatility in the peak areas, and overestimate the volatility in the trough areas. While, in real quantitative strategies, trading opportunities always happen in these areas. For example, in option trading, traders tend to sell option if the implies volatility (IV) \footnote{In financial mathematics, the implied volatility (IV) of an option contract is the value of the volatility of the underlying instrument which will return a theoretical value equal to the current market price of the said option when we input in an option pricing model, e.g., the Black-Scholes or Black-Scholes-Merton models \citep{merton1978returns}. } is higher than the predicted volatility (i.e., the proxy of realized volatility (RV)) during the peak area of the volatility sequence; if the predicted volatility is underestimated, the trading can lose money. 
On the other hand, if the IV is lower than the predicted volatility during the trough area of the volatility series, the traders will buy option; if the predicted volatility is overestimated, the strategy will make less money.

The emphasis of this paper is on addressing practical problems that arise in implementing ARCH family models and the SVR-GARCH model. In such settings, it is common knowledge that the overestimation/underestimation can be too large, leading to a lack of interpretability, and other issues. For these reasons, it is well motivated to develop an augmented method to constrain the predicted volatility closer to the realized volatility during the peak and trough areas and towards less ``backward eavesdropping". With this goal in mind, we propose the blending and augmented algorithm on traditional ARCH model methods.

\section{Augmented blending volatility forecasting}

\subsection{Parametric and semi-parametric volatility models}
Let $P_t$ be asset price at time $t$ where $t\in [1,2,\ldots, T]$, the return of the asset at time $t$ can be obtained by the following equation:
\begin{equation}
r_t = \ln \left(\frac{P_t}{P_{t-1}}\right), \gap \forall\,\, t\in[2,3,\ldots, T].
\end{equation}

\paragraph{ARCH} The ARCH model is simply a linear model having the following form for ARCH($p$):
\begin{equation}
	h_t = \alpha_0 + \sum_{i=1}^{p} \alpha_i h_{t-1} + \eta_t,
\end{equation}
where $h_t$ is the conditional variance at time $t$, and $\eta_t$ is the error term \citep{engle1982autoregressive}. The conditional variance is postulated to be a linear function of the past $p$ innovations.

\paragraph{GARCH} The GARCH model can capture volatility clustering and improves the ARCH model making it extensively used to model real financial time series problem \citep{bollerslev1986generalized}. The GARCH($p,q$) model is defined as follows:
\begin{equation}
\begin{aligned}
	r_t &= a+a_t,	\\
	a_t &= \sqrt{h_t} \cdot \epsilon_t, \gap \epsilon_t \sim i.i.d.(0,1), \\
	h_t &= \alpha_0  + \sum_{i=1}^{p} \beta_i h_{t-i}+ \sum_{i=1}^{q}\alpha_i  a_{t-i}^2,
\end{aligned}
\end{equation}
where the innovation $\epsilon_t$ is an independent identically distributed (i.i.d.) random variable with zero mean and unit variance and it is common to set $a=0$. 

\paragraph{EGARCH} \citet{nelson1991conditional} developed the EGARCH to model the skewness of financial returns whilst the variance is positive. The EGARCH($p,q$) model is defined as follows:
\begin{equation}
\begin{aligned} 
\ln(h_t) &= \alpha_0 + \sum_{i=1}^{p}\beta_i \ln(h_{t-i}) + \sum_{i=1}^{q}\alpha_i 
\frac{
|a_{t-i}| +\gamma_i a_{t-i}
}{ \sqrt{h_{t-i}}}
 \\
\end{aligned}
\end{equation}
where $\gamma_i$ is the asymmetric response parameter.

\paragraph{GJR} To capture asymmetric response of volatility, \citet{glosten1993relation} introduced the GJR($p,q$) model:
\begin{equation}
	h_t = \alpha_0 +
	\sum_{i=1}^{q}(\alpha_i + \gamma_i S_{t-i}^-) a_{t-i}^2 +  \sum_{i=1}^{p}\beta_i h_{t-i} ,
\end{equation}
where 
\begin{equation}
	S_{t-i}^- = 
	\left\{
	\begin{aligned}
		1, \gap a_{t-i} <0 ;\\
		0, \gap \text{otherwise},
	\end{aligned}
	\right.
\end{equation}
where the parameters are nonnegative.

\paragraph{SVR-GARCH} Machine learning aided method has been explored to forecast the volatility more precisely. \citet{bezerra2017volatility} showed the SVR-GARCH with mixture of Gaussian kernels can achieve better performance. In the SVR-GARCH model, the output variable is $h_t$ and the input vector is $\bx_t=[a_{t-1}^2,h_{t-1}]^\top$:
\begin{equation}
r_t = f(r_{t-1}) + a_t,
\end{equation}
where $f(\cdot)$ is the decision function estimated by SVR for the mean equation. Then, the variance is estimated by 
\begin{equation}\label{equation:svr-garch}
\widetilde{h}_t = g(\widetilde{h}_{t-1}, a_{t-1}^2),
\end{equation}
where $g(\cdot)$ is the decision function estimated by SVR again. 

\paragraph{Performance measure}
In real applications, the measurement of $h_t$ is not observed directly and \citet{andersen1998answering} showed the following realized volatility (RV) is closer to the theoretical volatility:
\begin{equation}
	\widetilde{h}_t = \frac{1}{5} \sum_{i=0}^{4}r_{t-i}^2.
\end{equation}
In the following development of our methods, we only consider this volatility proxy while other proxies may alter the results. However, this issue is beyond the scope of this paper since any volatility proxy is an imperfect estimator of the true conditional variance \citep{patton2011volatility}. To measure the proposed methods numerically in the next section, we use the root mean square error (RMSE) to evaluate the prediction performance which is given by 
\begin{equation}\label{equation:rmse}
\text{RMSE}(\by, \widehat{\by})=\sqrt{\frac{1}{T} \sum_{t=1}^{T} (y_t-\widehat{y}_t)^2},
\end{equation}
where $y_t$ denote the observation at time $t$ and $\widehat{y}_t$ represent the prediction of $y_t$.
Further, the mean absolute error (MAE) is also considered:
\begin{equation}\label{equation:mae}
	\text{MAE}(\by, \widehat{\by})={\frac{1}{T} \sum_{t=1}^{T} \big|y_t-\widehat{y}_t)\big|}.
\end{equation}
In all scenarios, smaller RMSE and MAE indicate better performance. 
Apart from the numerical measurement on the methods, we also highlight the pictorial behaviors of the predictions during the \textit{peaks} and \textit{troughs} of the predicted time series.

To further evaluate the significance of difference, we follow the DM test \citep{diebold1995comparing, harvey1997testing, diebold2015comparing}. Given the MAE losses of two predicted time series ($\bff, \bg$) with the true sequence $\by$, the two-sided DM test follows the following null and alternative hypothesis:
\begin{equation}
\begin{aligned}
\text{H}_0: \text{MAE}(\by, \bff) - \text{MAE}( \by, \bg)=0; \\
\text{H}_1: \text{MAE}(\by, \bff) - \text{MAE}(\by, \bg)\neq 0. \\
\end{aligned}
\end{equation}
When the null hypothesis is rejected, there is evidence the two predicted sequences are different. 
Let the \textit{loss-differential} $d_i$ be defined as the following absolute deviation:
\begin{equation}
d_i = |f_i - y_i| - |g_i - y_i|,
\end{equation}
where $f_i, g_i, y_i$ are $i$-th elements of $\bff, \bg$, $\by$ respectively. When the time series is of length $T$, define $N=T^{1/3}+1$, the DM statistics is calculated as follows:
\begin{equation}
\text{DM} = \frac{\overline{d}}{
\sqrt{
\left[\eta_0 + 2 \sum_{k=1}^{N-1} \eta_k\right]/T 
}
}\sim \mathcal{N}(0,1),
\end{equation}
where 
$\overline{d} = \frac{1}{T}\sum_{i=1}^{T} d_i, \eta_k = \frac{1}{T} \sum_{i=k+1}^{T} (d_i - \overline{d})(d_{i-k}  -\overline{d})$. Thus, there is significant difference between the two sequences if $|\text{DM}|$ is larger than the two-tailed critical value of the standard normal distribution (i.e., 1.96 if the significance value is selected to be 0.05).

\subsection{Augmented blending volatility forecasting}
\paragraph{Blending-ARCH (BARCH)} Applying the various autoregressive conditional heteroscedasticity family models on the real market return data, one should find the models can always overestimate the volatility. In real quantitative trading, this will result in losing trading opportunities. To overcome the drawback, we propose the blending-ARCH model. Suppose $h_t^1, h_t^2, \ldots, h_t^N$ are predictions of $N$ parametric volatility models at time $t$ (which can be either ARCH, GARCH, GJR, or EGARCH with different parameters). Then the \textit{uniform blending} of the results can be obtained by 
\begin{equation}
h_t = \frac{1}{N} \left(h_t^1+h_t^2+\ldots +h_t^N\right).
\end{equation}
However, since most of the parametric models will overestimate the volatility, a simple mean of the predictions can still overestimate it. A better proposal is the \textit{linear blending}:
\begin{equation}
	h_t =w_0+  w_1 \cdot h_t^1+w_1\cdot h_t^2+\ldots +w_N \cdot h_t^N,
\end{equation}
where the weight vector $\bw=[w_1, w_2, \ldots, w_N, w_0]^\top \in \real^{N+1}$ is set to be the same one for different time $t$ and is learned from ordinary least squares (OLS) in our case. Let $\bh=[\widetilde{h}_1, \widetilde{h}_2, \ldots, \widetilde{h}_T]^\top\in \real^T$ be the vector containing the realized volatility of each time $t\in [1,2,\ldots, T]$, $\bX$ be the matrix containing the prediction of different parametric models:
\begin{equation}\label{equation:data-matrix}
\bX = 
\begin{bmatrix}
 h_1^1 & h_1^2 & \ldots & h_1^N &1\\
 h_2^1 & h_2^2 & \ldots & h_2^N&1\\
\vdots & \vdots &\ddots &\vdots& \vdots\\
 h_T^1 & h_T^2 & \ldots & h_T^N&1\\
\end{bmatrix}\in \real^{T\times (N+1)}.
\end{equation}
Then OLS solution of $\bw$ can be obtained by 
\begin{equation}
\bw= (\bX^\top\bX)^{-1}\bX^\top\bh,
\end{equation}
where $\bw$ is also the projection of $\bh$ onto the column space of $\bX$ \citep{strang1993introduction, lu2021rigorous, lu2022matrix}. For each time $t\in [1,2,\ldots, T]$, we want to predict $h_t=\bx_t^\top\bw$ as closer to $\widetilde{h}_t$ as possible where $\bx_t^\top$ is the $t$-th row of $\bX$. Any other machine learning method can be applied to predict $h_t$:
\begin{equation}
	h_t = h(\bx_t),  \gap \forall \,\, t\in [1,2,\ldots, T].
\end{equation}
In our implementation, we employ the neural network approach to do the prediction, hence termed \textit{BARCH-NN} in the sequel. Since the number of features obtained from the blending method is 
not large, we apply a simple neural network structure with only fully connected layer (Appendix~\ref{appendix:-mpstructure}).

\paragraph{Time series correlation}
To do the feature selection, we want to compute the correlation of different time series data. The feature with high mean correlation with other features will be deleted. Given two vectors $\bx_1, \bx_2$, the cosine correlation is defined as:
\begin{equation}\label{equation:correlation}
\text{Cosine}(\bx_1, \bx_2)= \frac{\bx_1^\top \bx_2}{|\bx_1|\cdot|\bx_2|}
\end{equation}
Therefore, for the data matrix $\bX$ in Eq~\eqref{equation:data-matrix}, where each column is given by $\widehat{\bx}_i$ for all $i$ in [1, 3, \ldots, $N$], the $(i,j)$-th element of the feature correlation matrix is given by:
\begin{equation}\label{equation:correlation-matrix}
	\text{Correlation}(\bX)_{i,j}= \text{Cosine}(\widehat{\bx}_i, \widehat{\bx}_j),
\end{equation}
where $\text{Correlation}(\bX)$ is a matrix of shape $N\times N$ (the bias feature is deleted for the correlation evaluation). We will see, the feature selection based on this correlation matrix is important to improve the prediction performance and provide significant difference with other models in the sequel. 
While other correlation algorithms are also explored in our experiments, the difference is not that large, e.g., Pearson correlation, temporal-weighted correlation, and generalized correlation \citep{tulchinsky2019finding}. Though other algorithms exist to select the features, e.g., random forest \citep{anani2018comparison}, mixture model \citep{murphy2012machine, lu2017hyperprior}, these complex models seem not to provide better performance in our experiments while the computation may require extra resources.

\paragraph{Effective ratio}
\citet{kaufman2013trading, kaufman1995smarter} suggested replacing the ``weight" variable in the exponential moving average (EMA) formula with a constant based on the \textit{efficiency ratio} (ER). And the ER is shown to provide promising results for financial forecasting via classic quantitative strategies \cite{lu2022exploring} where the ER of the closing price are calculated to decide the trend of the asset. This indicator is designed to measure the \textit{strength of a trend}, defined within a range from -1.0 to +1.0 where the larger magnitude indicates a larger upward or downward trend. Instead of calculating the ER of the closing price, we want to calculate the ER of the volatility series. Given the window size $M$, it is calculated with a simple formula:
\noindent
\begin{equation}
\begin{aligned}
	e_t(M)  &= \frac{s_t}{n_t}= \frac{h_{t-1} - h_{t-1-M}}{\sum_{i=1}^{M} |h_{t-i} - h_{t-1-i}|}\\
&= \frac{\text{Total volatility change for a period}}{\text{Sum of absolute volatility change for each bar}},
\end{aligned}
\end{equation}
where $e_t(M)$ is the ER at time $t$. We carefully notice that the ER at time $t$ is based on $h_{t-1}, h_{t-2}, \ldots, h_{t-1-M}$ to avoid a forward bias since we do not know $h_t$ at time $t$.
At a strong trend (i.e., the input volatility is moving in a certain direction, up or down) the ER will tend to 1 in absolute value; if there is no directed movement, it will be a little more than 0. 

\paragraph{Augmented BARCH (aBARCH) and augmented SVR-GARCH (aSVR-GARCH)} The aBARCH method simply adds the effective ratio to the predicted results of bARCH:
\begin{equation}
h_t = w_0+  w_1 \cdot h_t^1+w_1\cdot h_t^2+\ldots +w_N \cdot h_t^N + \sigma \cdot e_t(M),
\end{equation}
where the window size $M$ and deviation $\sigma$ are hyperperameters that can be tuned by cross validation (CV).
We carefully notice that, when $M$ is larger, $ e_t(M)$ will usually tend to 0, thus $\sigma \cdot e_t(M)$ tend to 0; and when $\sigma\rightarrow 0$, $\sigma \cdot e_t(M)$ will also tend to 0, which reduces to the trivial case. 
Similarly, the aSVR-GARCH can be obtained based on Eq.~\eqref{equation:svr-garch}: 
\begin{equation}
	\widetilde{h}_t = g(\widetilde{h}_{t-1}, a_{t-1}^2)+ \sigma\cdot e_t(M).
\end{equation}
In practice, $M=15$ and $\sigma=0.1$ can be a good candidate since it only puts a small amount of value to the augmented prediction.

\begin{table}[!h]
	\begin{tabular}{lll}
		\hline
		Statstics    & \gap SH300 return & \gap   S\&P500 return \\ \hline
		Observations & \gap 2956         & \gap 2778       \\
		Mean         & \gap 0.00017      &  \gap -0.00041      \\
		Std.dev      & \gap 0.01429      & \gap  0.01086      \\
		Median       & \gap 0.00034      & \gap  -0.00069      \\
		Kurtosis     & \gap 4.42880      & \gap  19.72931      \\
		Skewness     & \gap -0.53949     & \gap  1.23705      \\
		Maximum      & \gap 0.06715      & \gap  0.13616      \\
		Minimum      & \gap -0.08748     & \gap  -0.08578      \\ \hline
	\end{tabular}
	\caption{Descriptive statistics for daily returns.}
	\label{fig:table-data-info}
\end{table}

\begin{figure*}[!h]
\centering
\subfigure[GARCH-N(1,1)
 ]{\includegraphics[width=0.24\textwidth, ]{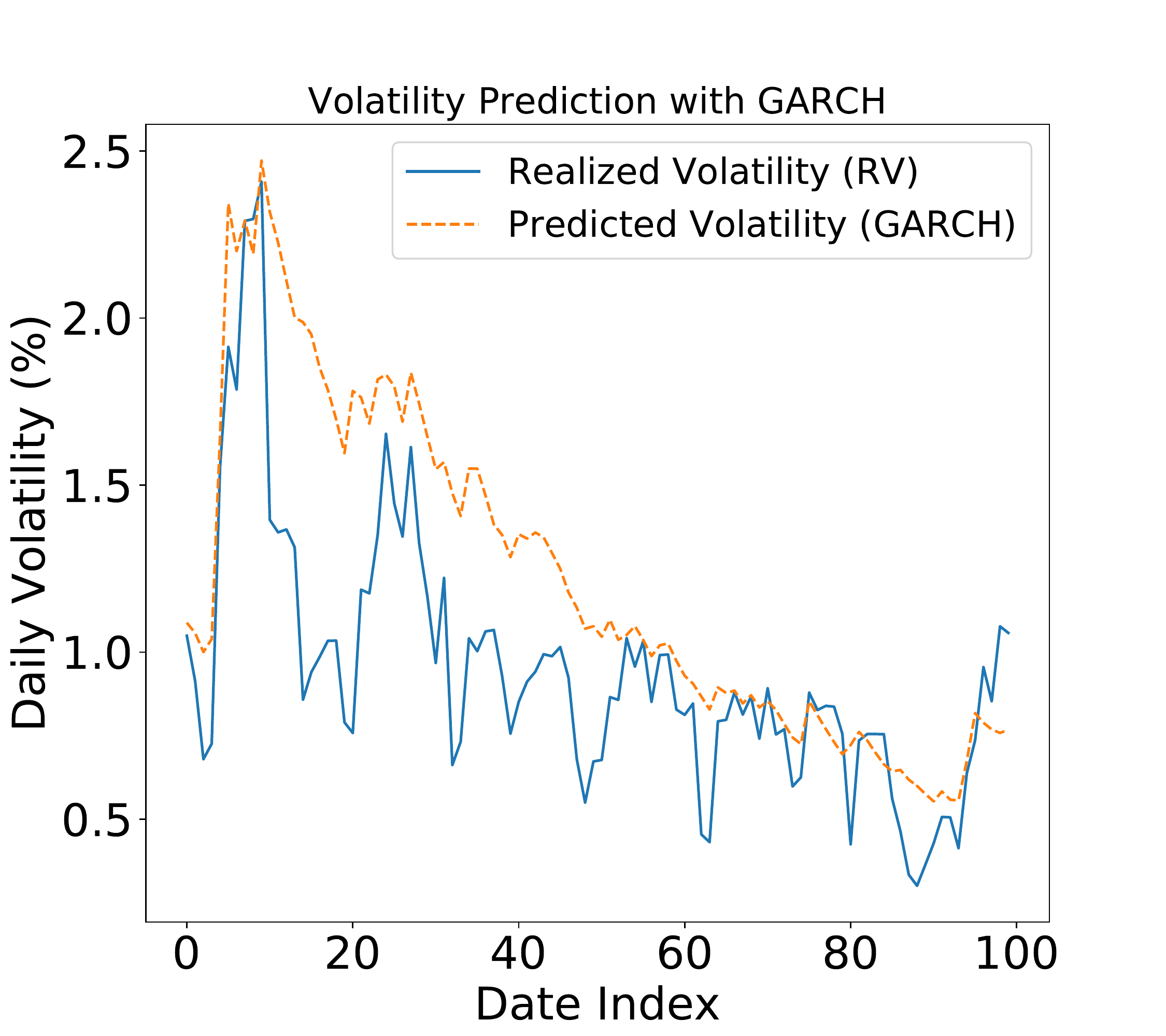} \label{fig:typical_garch}}
\hfill
\subfigure[SVR-GARCH
]{\includegraphics[width=0.24\textwidth]{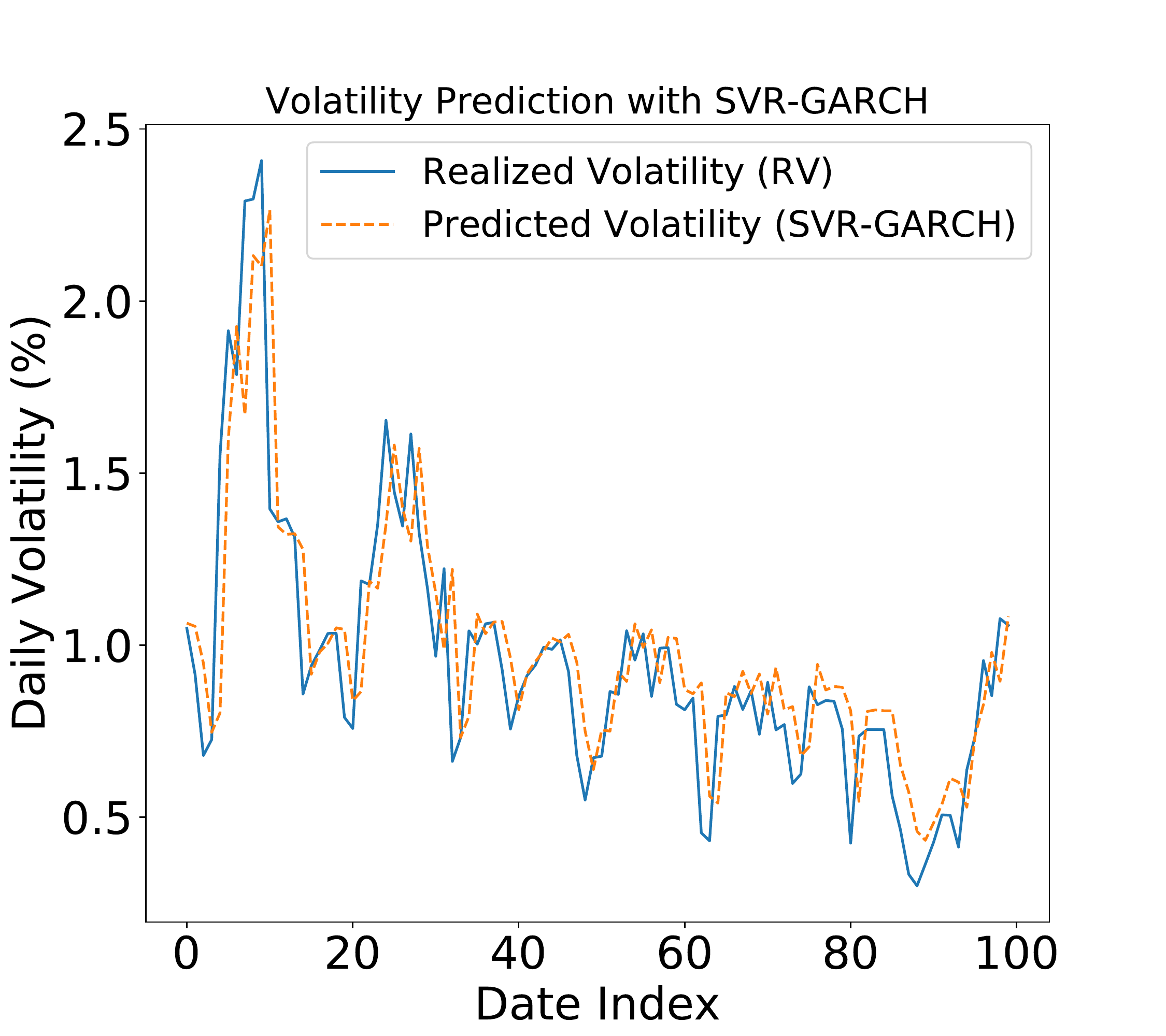} \label{fig:typical_svrgarch}}
\hfill
\subfigure[BARCH(75)
]{\includegraphics[width=0.24\textwidth]{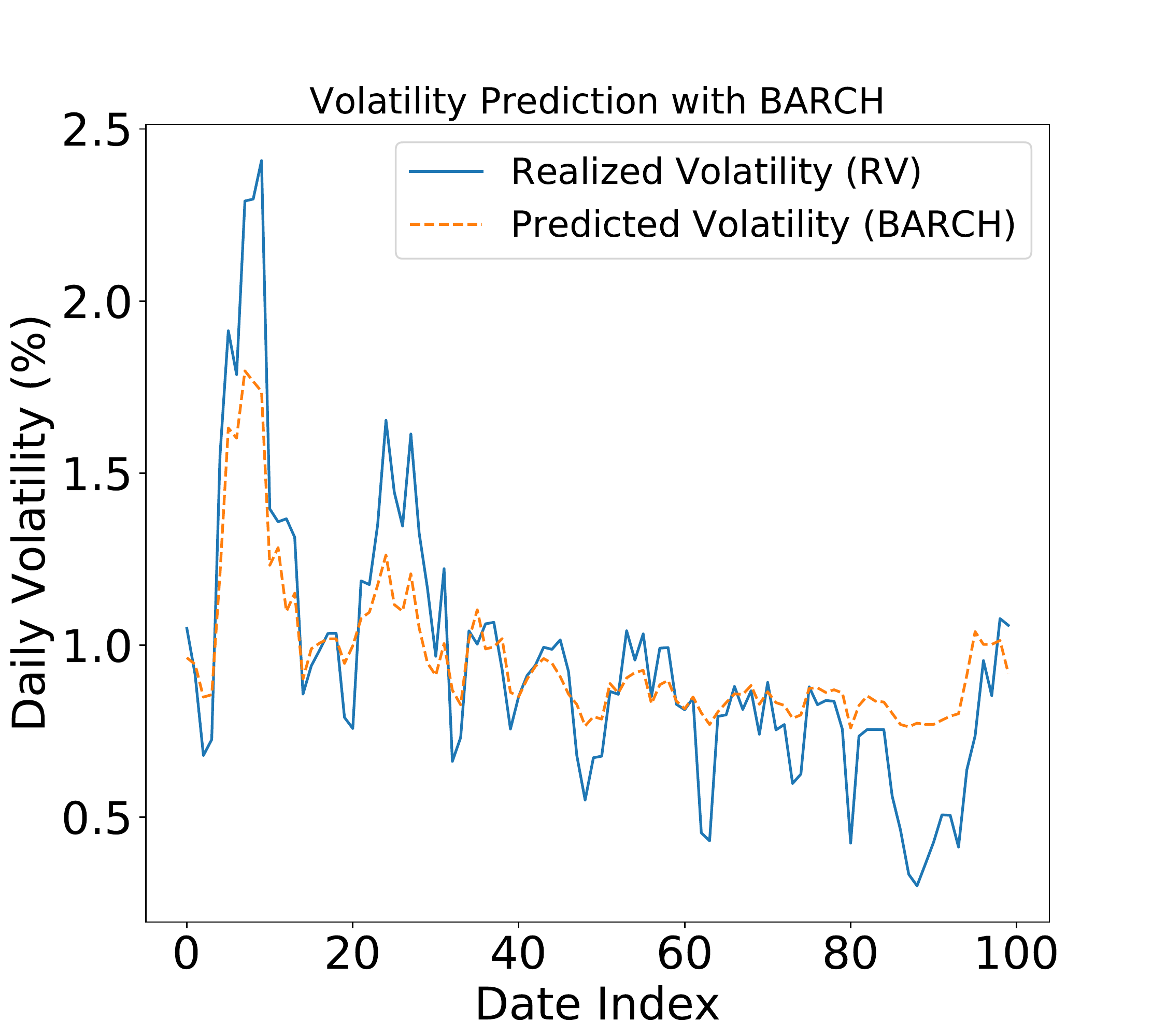} \label{fig:typical_barch}}
\hfill
\subfigure[BARCH-NN(55)
]{\includegraphics[width=0.24\textwidth]{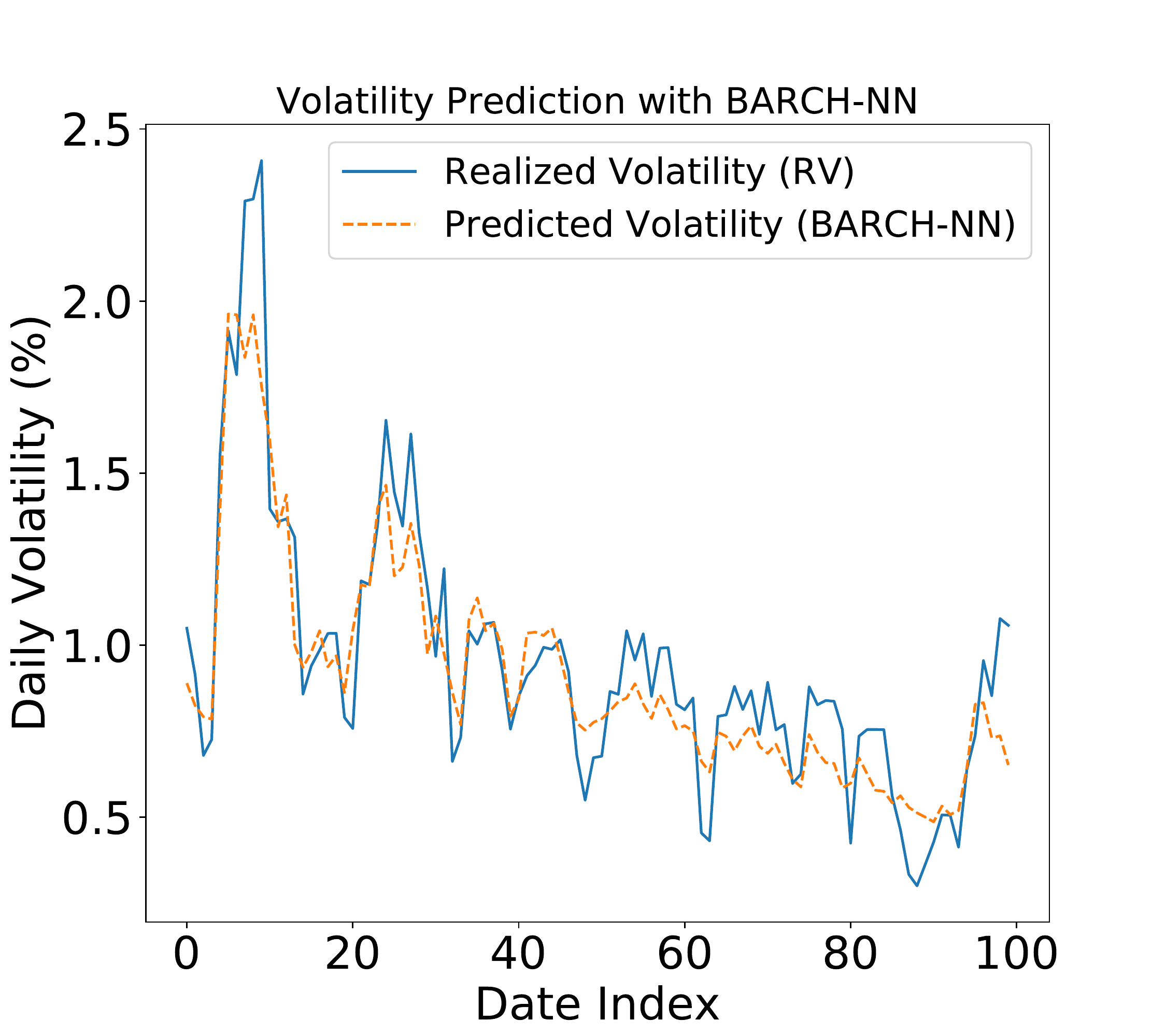} \label{fig:typical_barch-nn}}
\center
\caption{\textbf{SH300}: Pictorial behaviors of GARCH, SVR-GARCH, BARCH models on SH300 data. We can observe the ``backward eavesdropping" problem in SVR-GARCH model in Figure~\ref{fig:typical_svrgarch}. The BARCH(75) and BARCH-NN(55) models reduce the ``backward eavesdropping" problem to some extent. Though the BARCH-NN(55) may look ``worse" than SVR-GARCH from the picture at first glance, the RMSE  and MAE performances of the former one are better.}
\label{fig:typical-behavior}
\end{figure*}

\begin{figure*}[!h]
	\center
	\subfigure[BARCH(75) vs aBARCH(75)]{\includegraphics[width=0.32\textwidth]{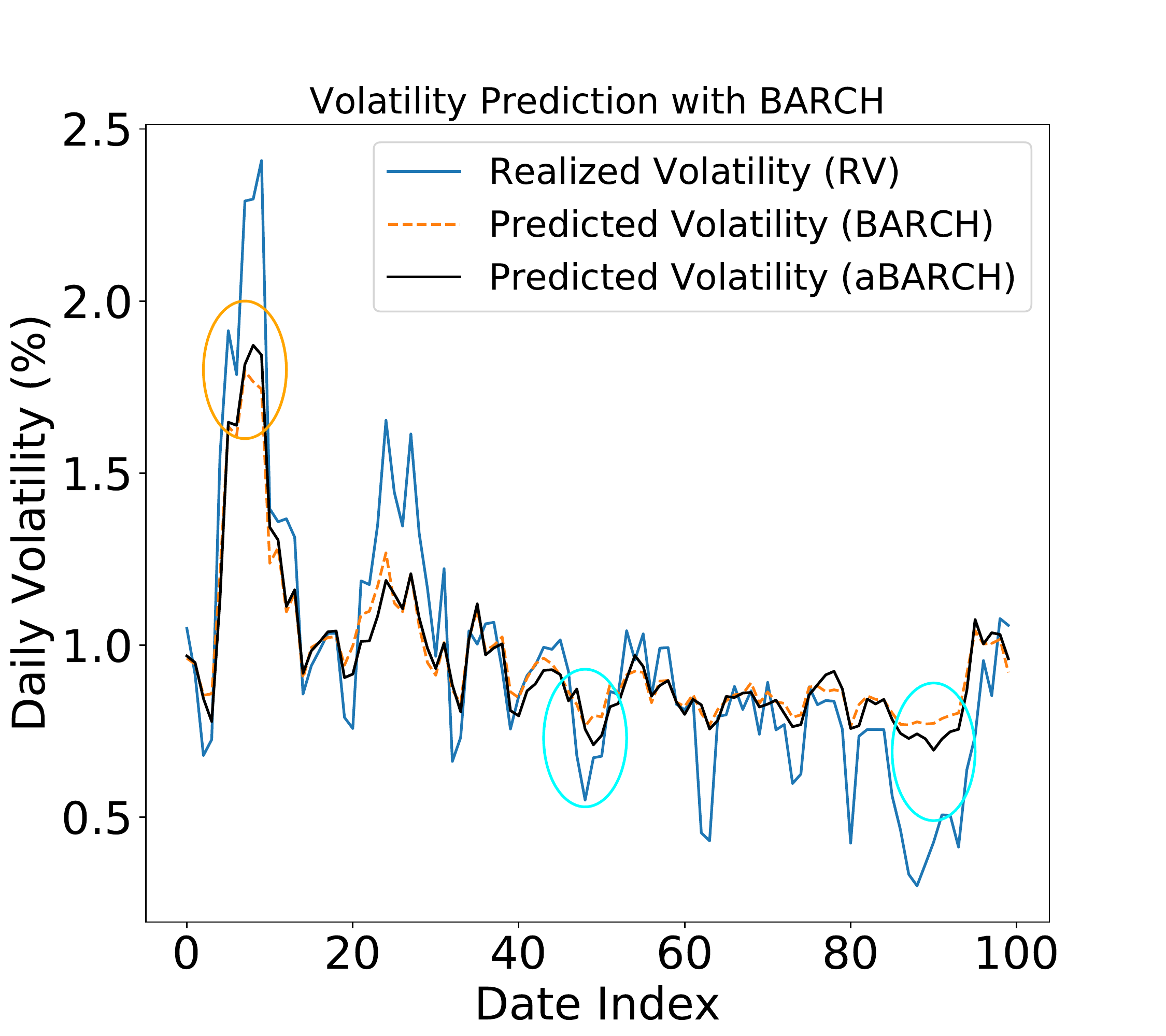} \label{fig:sh300-typical-er1}}
	\hfill
	\subfigure[BARCH-NN(55) vs aBARCH-NN(55)]{\includegraphics[width=0.32\textwidth]{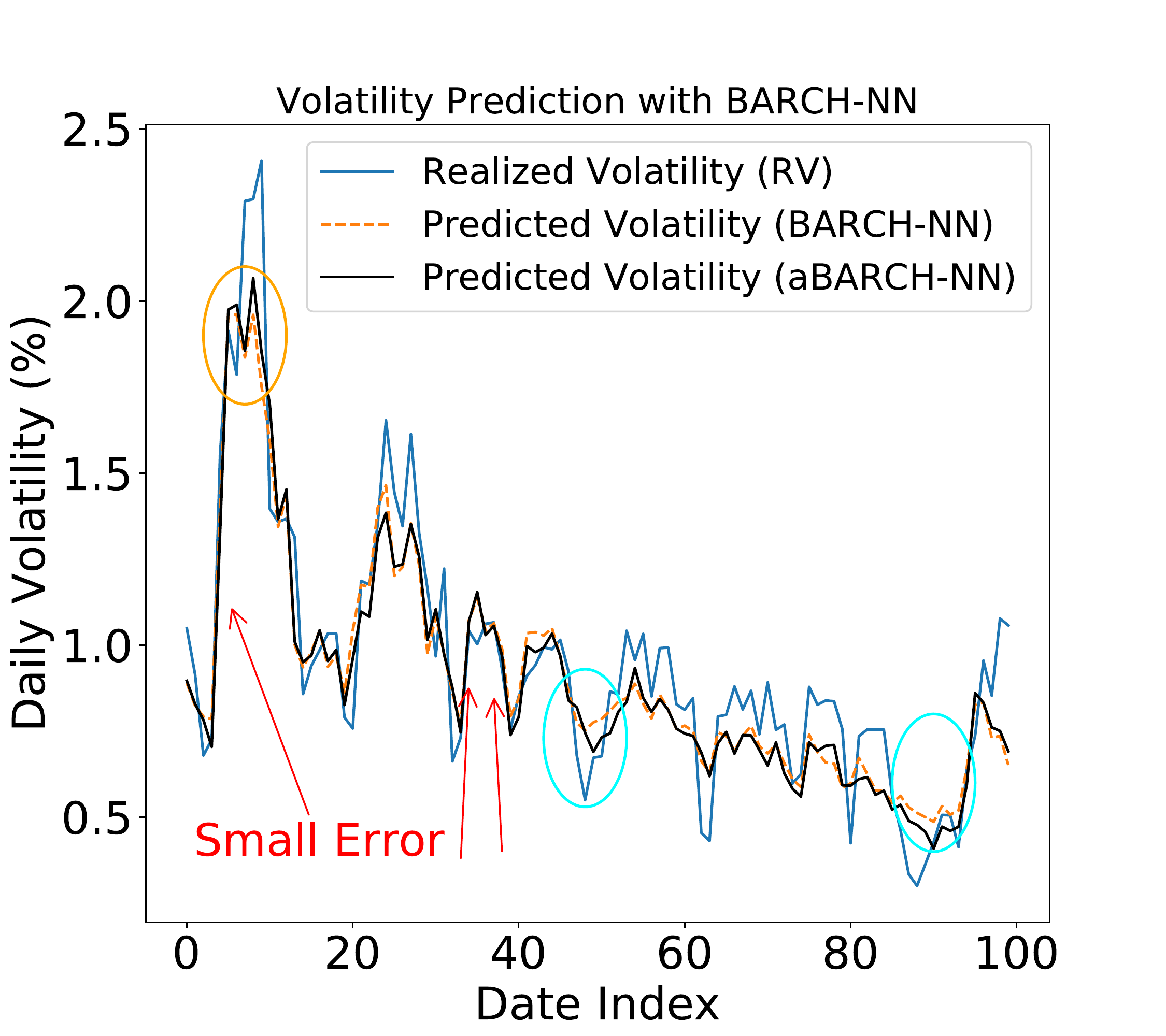} \label{fig:sh300-typical-er2}}
	\hfill
	\subfigure[SVR-GARCH vs aSVR-GARCH]{\includegraphics[width=0.32\textwidth]{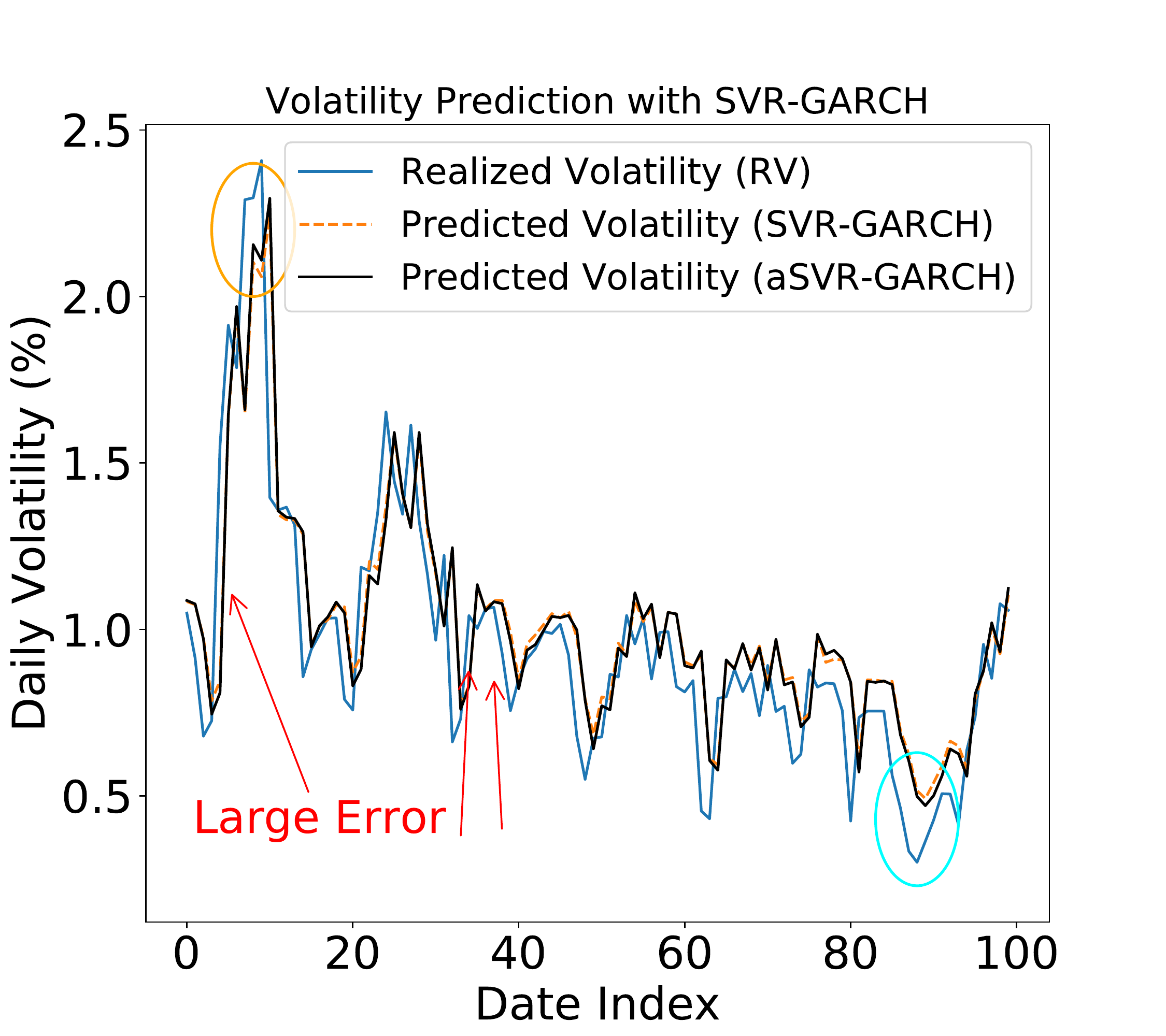} \label{fig:sh300-typical-er3}}
	\center
	\caption{\textbf{SH300}: Demonstration of the augmented method on each model for reducing underestimation (orange cycles) and overestimation (cyan cycles) on SH300 data. Compare Figure~\ref{fig:sh300-typical-er2} and Figure~\ref{fig:sh300-typical-er3} shows the BARCH-NN(55) and aBARCH-NN(55) models perform better in the rising edge and descending edge of the volatility than the SVR-GARCH model.}
	\label{fig:sh300-typical-er}
\end{figure*}

\begin{figure*}[!h]
	\center
	\subfigure[BARCH(35) vs aBARCH(35)]{\includegraphics[width=0.32\textwidth]{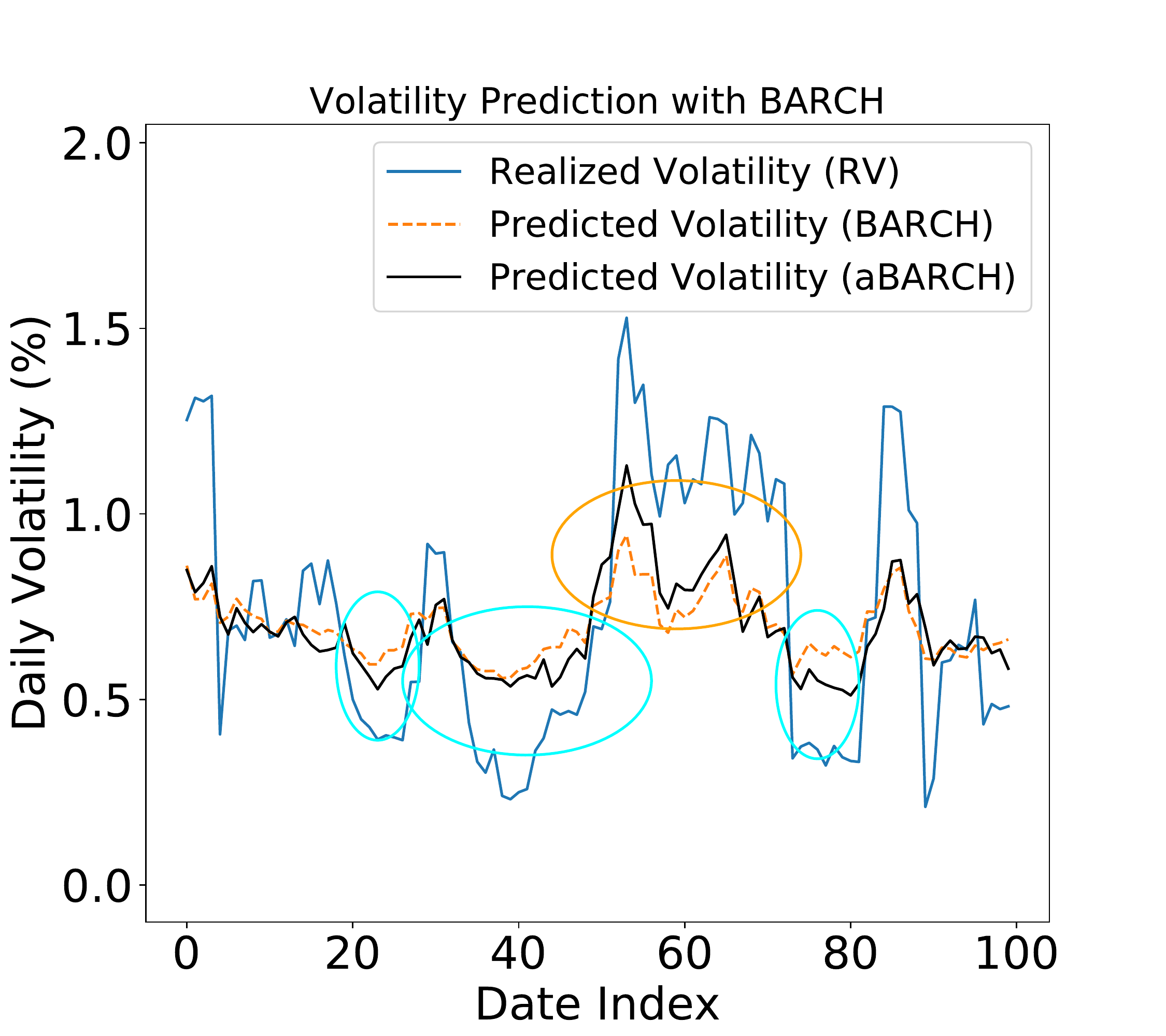} \label{fig:sp500-typical-er1}}
	\hfill
	\subfigure[BARCH-NN(15) vs aBARCH-NN(15)]{\includegraphics[width=0.32\textwidth]{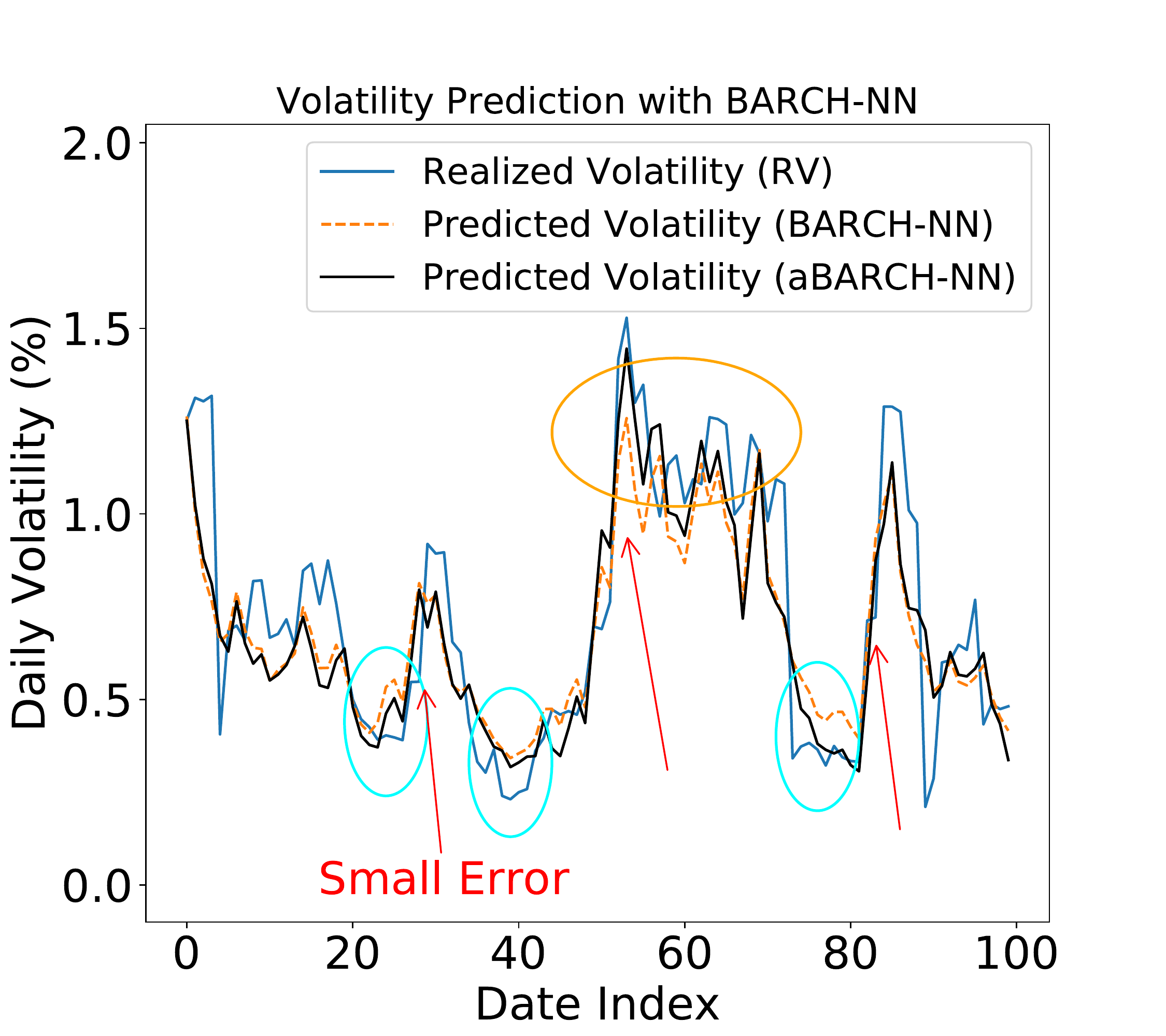} \label{fig:sp500-typical-e2}}
	\hfill
	\subfigure[SVR-GARCH vs aSVR-GARCH]{\includegraphics[width=0.32\textwidth]{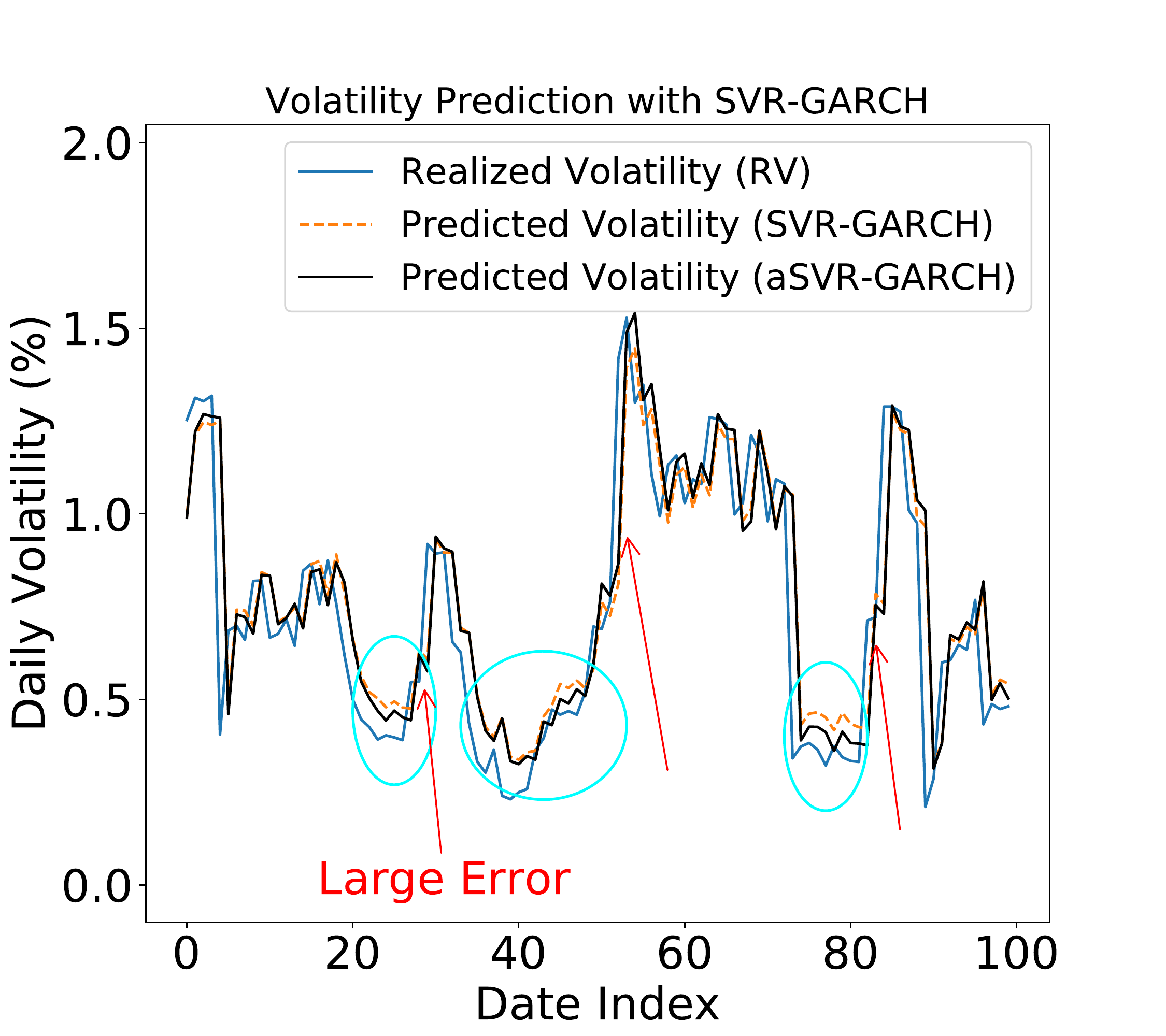} \label{fig:sp500-typical-er3}}
	\center
	\caption{\textbf{S\&P500}: Demonstration of the augmented method on each model for reducing underestimation (orange cycles) and overestimation (cyan cycles) on S\&P500 data. Compare Figure~\ref{fig:sp500-typical-e2} and Figure~\ref{fig:sp500-typical-er3} shows the BARCH-NN(15) and aBARCH-NN(15) models perform better in the rising edge and descending edge of the volatility than the SVR-GARCH model.}
	\label{fig:sp500-typical-er}
\end{figure*}

\section{Experiments}

To evaluate the strategy and demonstrate the main advantages of the proposed BARCH, BARCH-NN, and augmented methods, real market data sets are used. In a wide range of scenarios across various feature settings, BARCH-NN and aBARCH-NN increase the prediction ability in terms of RMSE and MAE measurements on out-of-sample prediction; and augmented method along can reduce overestimation and underestimation in peak and trough areas.

We obtain the S\&P500 index data, which
is a market-capitalization-weighted index of 500 leading publicly traded companies in the
U.S. with a time period of 11 years (between Jan. 3, 2011 and
Jan. 14, 2022).
The S\&P500 index uses a market-cap weighting method, giving a higher
percentage allocation to companies with the larger market-capitalizations.
Further, we obtain the SH300 index data, a Chinese alternative for the S\&P500
(similar  to S\&P500, but still has a large difference, which is a market-capitalization weighted index of 300 leading publicly traded companies in China), with a time period of 12 years (between Jan. 4, 2010 and Mar. 7, 2022). 
Table~\ref{fig:table-data-info} shows the summary of the descriptive statistics for the SH300 and S\&P500 over the whole period we have obtained.
For each of ARCH, GARCH, EGARCH, GJR models, four different distributions for innovations are considered, i.e., the normal , the Student's t, the skewed Student's t, and the generalized error distribution (GED). For each innovation, the parameters are tuned by selecting the smallest Bayesian information criterion (BIC) and the best parameters will be represented in the model name (e.g., GARCH-N(1,1) in Table~\ref{fig:sh300_out-of-sample} or Table~\ref{fig:sp500_out-of-sample}). The goodness of fit for SH300 and S\&P500 are shown in Table~\ref{fig:sh300_goodness} and Table~\ref{fig:sp500_goodness} respectively in Appendix~\ref{appendix:goodness-fit}. For the SVR-GARCH model, the parameters are tuned by cross validation.

In all scenarios, we allocate the first 2456 or 2278 observations (for SH300 and S\&P500 respectively) for training, the next 252 observations for validation, and the last 252 observations for testing \footnote{252 days is usually the total number of trading days per year in the U.S. and is widely used in research.}.  
The training and validation samples constitute the in-sample data, and the test samples constitute the out-of-sample data.

In all experiments,  the features for BARCH or BARCH-NN are obtained with a set of different parameters on ARCH, GARCH, EGARCH, or GJR models (say different $p, q$, or distributions for innovations). The methods will be denoted as \textit{BARCH($K$)} or \textit{BARCH-NN($K$)} if the number of features is $K$. In all scenarios, the features are selected randomly from the total feature sets (90 features totally in our experiments). 

\begin{figure}[!h]
\center
\subfigure[Heatmap of SH300]{\includegraphics[width=0.4\textwidth]{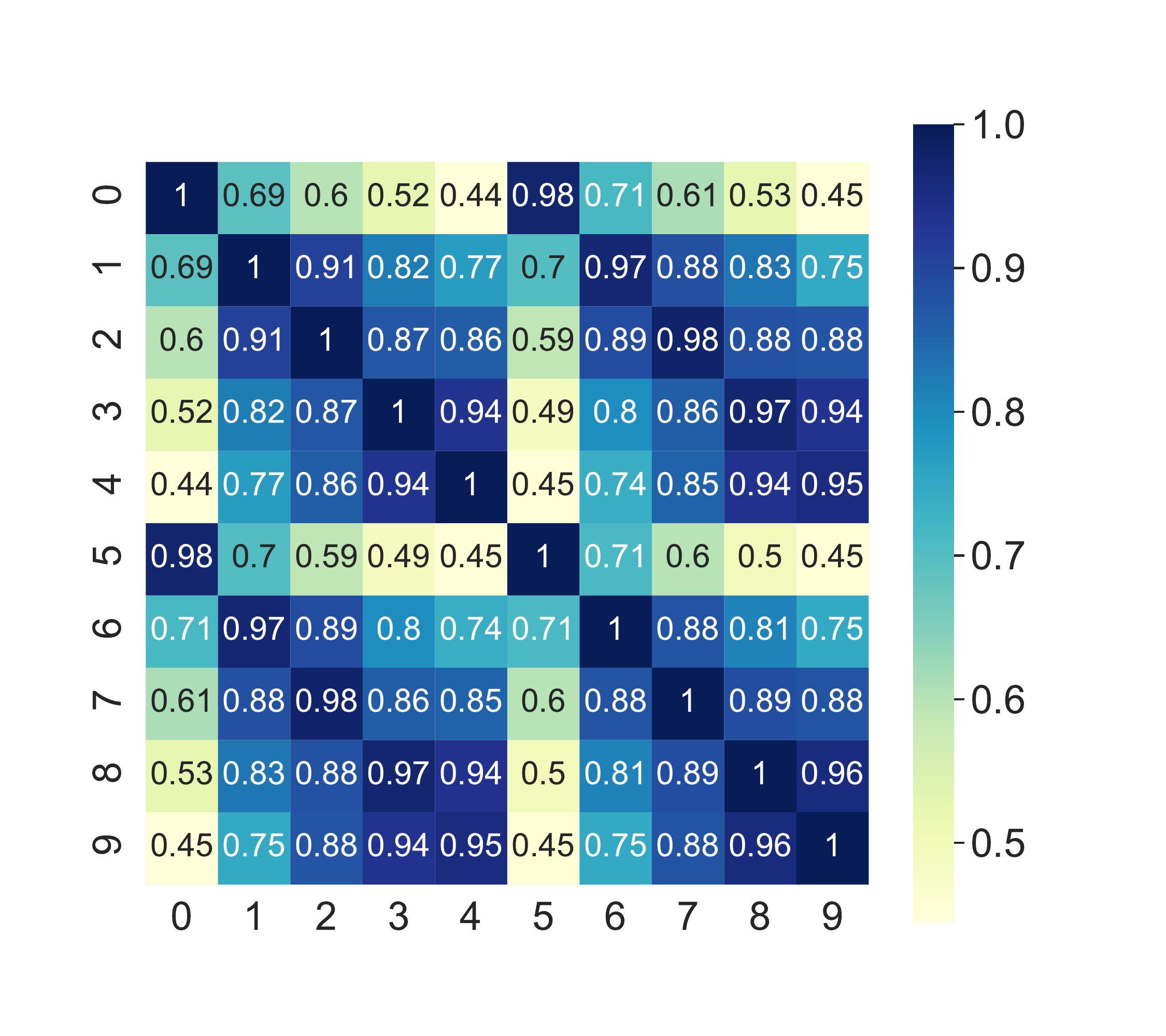} \label{fig:heatmap-sh300}}
\hfill
\subfigure[Heatmap of S\&P500]{\includegraphics[width=0.4\textwidth]{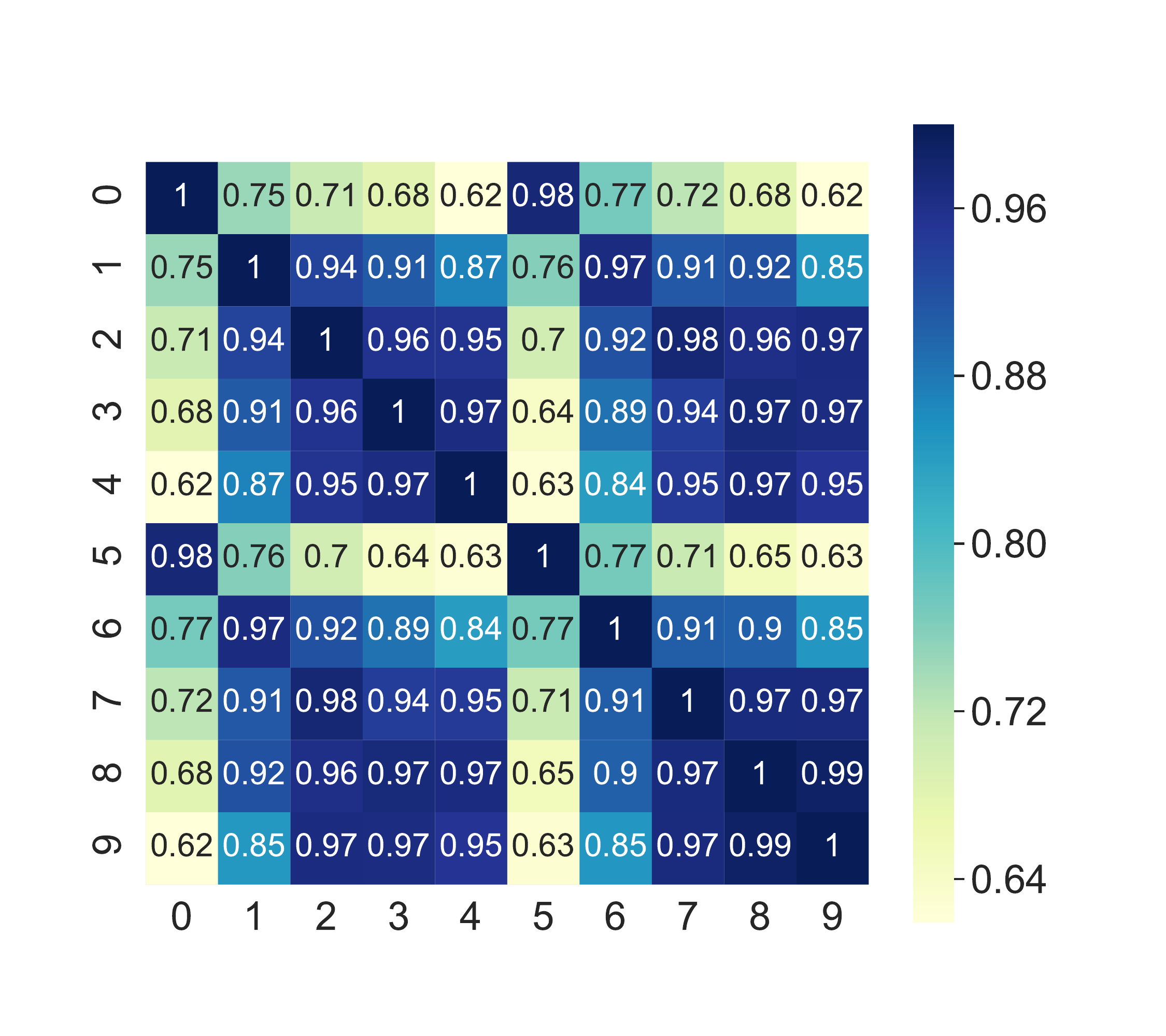} \label{fig:heatmap-sp500}}
\center
\caption{The heatmaps of the correlation among selected features for SH300 and S\&P500 data sets.}
\label{fig:heatmapsr}
\end{figure}

Based on Eq.~\eqref{equation:correlation-matrix}, the correlation matrix of the training features is obtained, a matrix of shape $\real^{90\times 90}$ in our case. Then any feature with a mean correlation smaller than 0.9 is selected afterwards. For both the SH300 and the S\&P500 data sets, 10 features are finally selected. The heatmaps of the selected 10 features are shown in Figure~\ref{fig:heatmapsr} for the two data sets. While the correlated based feature selection procedure for the models will be denoted as \textit{BARCH(CO)} or \textit{BARCH-NN(CO)} in the experiments.

\begin{figure}[!h]
	\center
	\subfigure[\textbf{SH300}: BARCH-NN(CO) vs aBARCH-NN(CO)]{\includegraphics[width=0.4\textwidth]{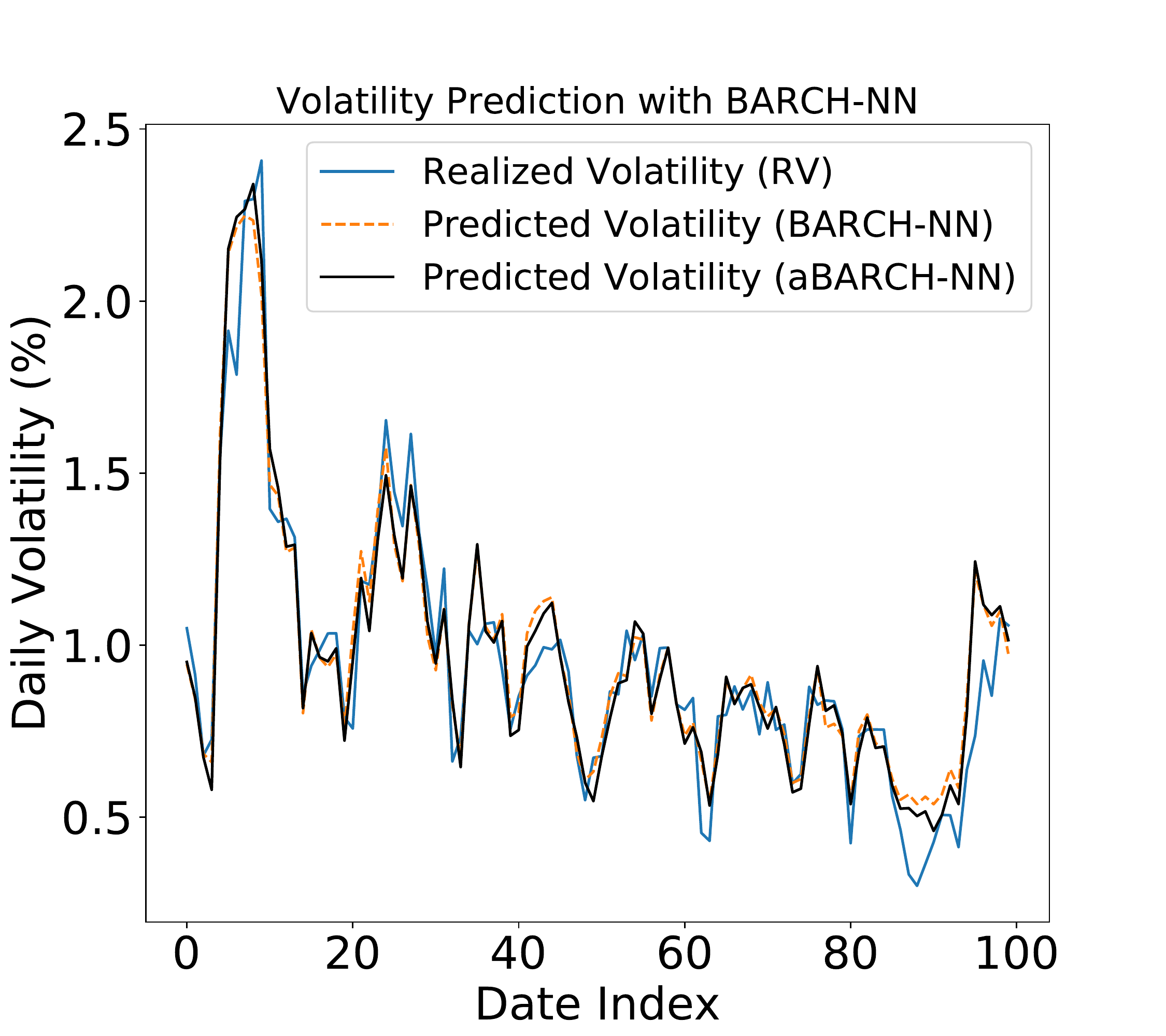} \label{fig:corr-sh300}}
	\hfill
	\subfigure[\textbf{S\&P500}: BARCH-NN(CO) vs aBARCH-NN(CO)]{\includegraphics[width=0.4\textwidth]{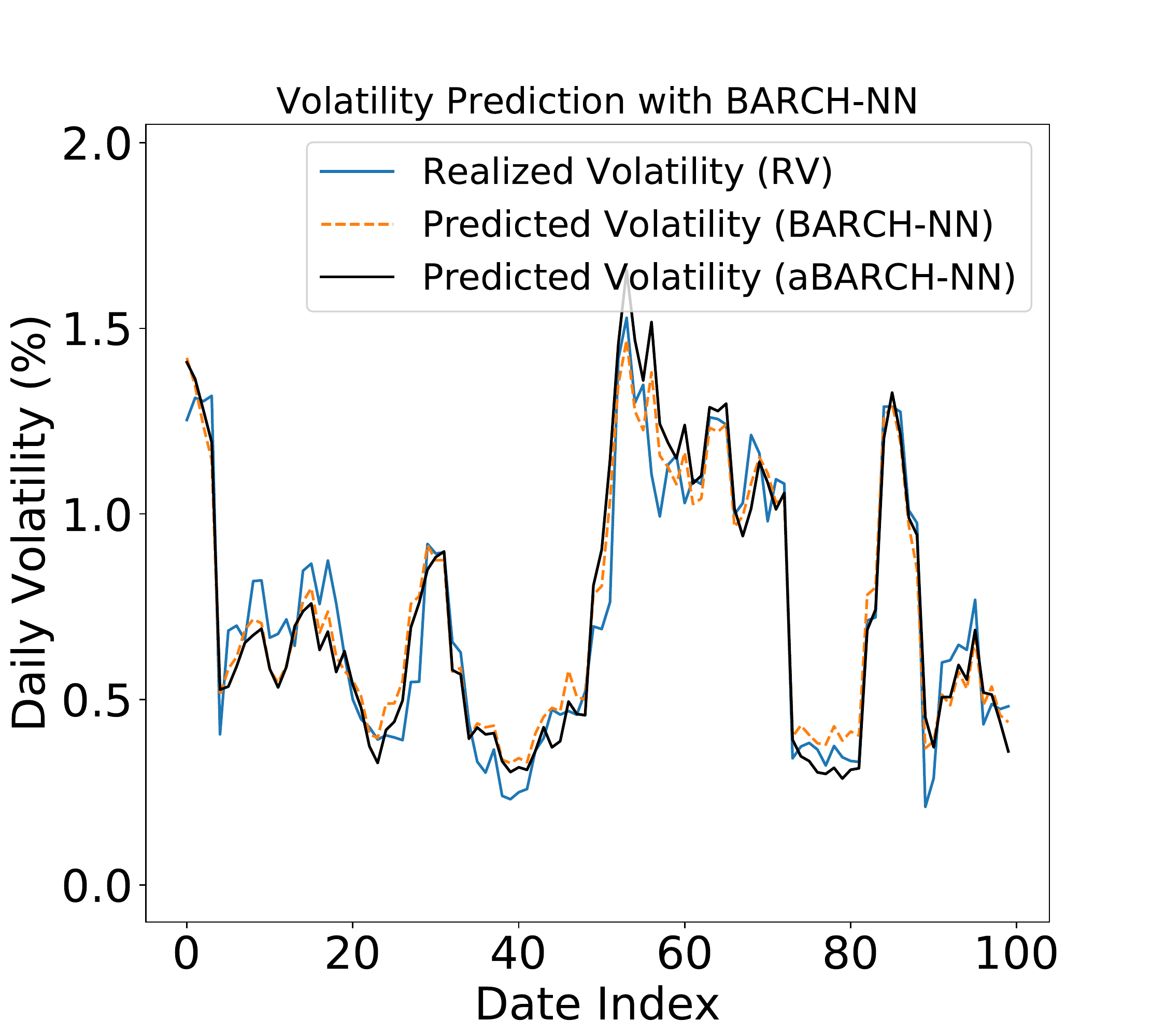} \label{fig:corr-sp500}}
	\center
	\caption{Best out-of-sample performances on SH300 and S\&P500 data sets.}
	\label{fig:correlated-results}
\end{figure}

\begin{table}[!h]
	\small
\begin{tabular}{lll}
\hline
Model     &  RMSE ($\times10^{-3}$)         & MAE ($\times10^{-3}$) \\ \hline
\underline{Eavesdrop} & \underline{2.279} & \underline{1.484}\\
BARCH(5)        &    2.593 & 1.917      \\
BARCH(15)        &     2.305 & 1.753       \\
BARCH(35)        &   2.287 & 1.730      \\
BARCH(55)        &  2.306             &  1.740       \\
BARCH(75)        &   2.302 & 1.737      \\
BARCH(CO)        &   2.217 & 1.724    \\
BARCH-NN(5)    &  2.226 & 1.703     \\
BARCH-NN(15)    &  {2.045}      &  {1.604}       \\
BARCH-NN(35)    &  {1.910}      &  {1.524}       \\
BARCH-NN(55)    &  \textbf{1.785}      &  \textbf{1.419}       \\
BARCH-NN(75)    &  {2.045}      &  {1.582}       \\
BARCH-NN(CO)    &  \underline{\textbf{1.289 }  }    &  \underline{\textbf{0.959} }      \\
aBARCH(5)       &  2.453& 1.803    \\
aBARCH(15)       &  2.170& 1.633    \\
aBARCH(35)       &  2.173& 1.641 \\
aBARCH(55)       &  2.196      &     1.656    \\
aBARCH(75)       &   2.192& 1.652  \\
aBARCH(CO)       &   2.121& 1.621  \\
aBARCH-NN(5)       &  2.086& 1.600      \\
aBARCH-NN(15)       &  {1.938}      &   {1.517 }      \\
aBARCH-NN(35)       &  {1.848}      &   {1.512 }      \\
aBARCH-NN(55)       &  \textbf{1.689}      &   \textbf{1.359}      \\
aBARCH-NN(75)       &  1.904     &   1.455     \\
aBARCH-NN(CO)       & \underline{\textbf{1.295}}      &   \underline{\textbf{0.948}  }    \\
SVR-GARCH    &  2.223      &     1.516    \\
aSVR-GARCH     &  2.421     &    1.649     \\
ARCH-N(14)      &  7.261   &  4.796      \\
ARCH-t(11)     &   8.094     & 5.524     \\ 
ARCH-st(11)     &  8.106     &  5.532       \\ 
ARCH-G(10)     &  7.251    &   4.847   \\ 
GARCH-N(1,1)     &  5.617    &   4.041     \\ 
GARCH-t(1,1)     &   5.693   &   4.241     \\ 
GARCH-st(1,1)     &  5.699    &   4.247     \\ 
GARCH-G(1,1)    &   5.356   &    3.936    \\ 
EGARCH-N(1,1)      & 6.932     &   5.167      \\ 
EGARCH-t(1,1)      & 7.108     &   5.441     \\ 
EGARCH-st(1,1)     & 7.118     &   5.453     \\ 
EGARCH-G(1,1)     &   6.699    &    5.058    \\ 
GJR-N(1,1)     &   5.773   &   4.126    \\ 
GJR-t(1,1)     &    6.526   &   4.668     \\ 
GJR-st(2,1)     &  6.524    &   4.671     \\ 
GJR-G(1,1)     &  5.702    &    4.124    \\ 
\hline
\end{tabular}
\caption{\textbf{SH300}: Out-of-sample evaluation on SH300 data.}
\label{fig:sh300_out-of-sample}
\end{table}

\begin{table}[!h]
\small
\begin{tabular}{lll}
\hline
Model     &  \gap DM Statistics          &\gap\gap P-value \\ \hline
BARCH-NN(15) &\gap -0.061 &\gap\gap 0.9516 \\
BARCH-NN(35) &\gap 0.822 &\gap\gap 0.4119 \\
BARCH-NN(55) &\gap {1.971} & \gap\gap 0.0499 \\
BARCH-NN(75) &\gap 0.175 &\gap\gap 0.8610 \\
BARCH-NN(CO) &\gap \textbf{5.916} & \gap\gap 0.0000 \\
aBARCH-NN(15) &\gap {2.464} & \gap\gap 0.0144 \\
aBARCH-NN(35) &\gap 0.991 &\gap\gap 0.3226 \\
aBARCH-NN(55) &\gap {2.762} & \gap\gap 0.0062 \\
aBARCH-NN(75) &\gap 1.571 &\gap\gap 0.1174 \\
aBARCH-NN(CO) &\gap \textbf{6.249} & \gap\gap 0.0000 \\
aSVR-GARCH &\gap -0.570 &\gap\gap 0.5690 \\
\hline
\end{tabular}
\caption{\textbf{SH300}: DM test on SH300 data (benchmark is the SVR-GARCH result).}
\label{fig:sh300_out-of-sample-mdtest}
\end{table}

\begin{table}[!h]
	\small
	\begin{tabular}{lll}
		\hline
		Model        &  RMSE ($\times10^{-3}$)         & MAE ($\times10^{-3}$) \\ \hline
		\underline{Eavesdrop} & \underline{3.700} & \underline{2.241} \\
		BARCH(5)        &  5.274 & 3.967     \\
		BARCH(15)        & 5.080 & 3.906     \\
		BARCH(35)        &  4.739 & 3.699     \\
		BARCH(55)        &  4.739            &    3.715     \\
		BARCH(75)        &  4.753 & 3.723     \\
		BARCH(CO)        &  5.309 & 4.161   \\
		BARCH-NN(5) &  3.728    &   2.502    \\
		BARCH-NN(15) &  \textbf{3.018 }   &   {2.287}    \\
		BARCH-NN(35) &  3.237    &    \textbf{2.138}    \\
		BARCH-NN(55) &  {3.242}      &  {2.306}       \\
		BARCH-NN(75) &  3.559    &   2.485   \\
		BARCH-NN(CO) &  \underline{\textbf{2.262  }}   &  \underline{\textbf{1.580}}    \\
		aBARCH(5)        &   5.079& 3.792    \\
		aBARCH(15)        &    4.889& 3.736      \\
		aBARCH(35)        &   4.560& 3.530    \\
		aBARCH(55)       &  4.563            &    3.556    \\
		aBARCH(75)        &   4.578& 3.563        \\
		aBARCH(CO)        &   5.137& 3.995       \\
		aBARCH-NN(5) &     3.617& 2.463    \\
		aBARCH-NN(15) &  \textbf{2.950}     &  {2.189}      \\
		aBARCH-NN(35) & 3.117     &   \textbf{2.064}     \\
		aBARCH-NN(55)&  {3.121}      &   {2.238}      \\
		aBARCH-NN(75) &  3.405    &    2.345    \\
		aBARCH-NN(CO)        &  \underline{\textbf{2.203}}  & \underline{\textbf{1.568}}         \\
		SVR-GARCH    &   3.542       &   2.218     \\
		aSVR-GARCH   &   3.586       &   2.253     \\
		ARCH-N(6)      &  20.811     &    8.626     \\
		ARCH-t(6)     &  22.426     &    9.272     \\ 
		ARCH-st(6)     &  22.296     &    9.222     \\ 
		ARCH-G(6)     &  21.631     &    8.931     \\ 
		GARCH-N(2,1)     &  20.484    &  8.913     \\ 
		GARCH-t(2,1)     &    21.872  &   9.468     \\ 
		GARCH-st(2,1)     &  21.717    &  9.417      \\ 
		GARCH-G(2,1)     &  21.181    &   9.179     \\ 
		EGARCH-N(2,1)     &   11.691   &    5.833    \\ 
		EGARCH-t(2,1)     &  12.805    &   6.387     \\ 
		EGARCH-st(2,1)     &  12.584    &  6.890      \\ 
		EGARCH-G(2,1)     &  12.234    & 6.080       \\ 
		GJR-N(1,1)     &   12.659   &    7.684    \\ 
		GJR-t(1,1)     &  13.795    &  8.241      \\ 
		GJR-st(1,1)     &  13.777    &  8.239      \\ 
		GJR-G(1,1)     &  13.344    &   8.009     \\ 
		\hline
	\end{tabular}
	\caption{\textbf{S\&P500}: Out-of-sample evaluation on S\&P500 data.}
	\label{fig:sp500_out-of-sample}
\end{table}
\begin{table}[!h]
	\small
	\begin{tabular}{lll}
		\hline
		Model     &\gap  DM Statistics          &\gap\gap P-value \\ \hline
BARCH-NN(15) &\gap -0.216 &\gap\gap 0.8290 \\
BARCH-NN(35) &\gap 0.802 &\gap\gap 0.4234 \\
BARCH-NN(55) &\gap {-3.706} & \gap\gap 0.0003 \\
BARCH-NN(75) &\gap -1.915 &\gap\gap 0.0567 \\
BARCH-NN(CO) &\gap \textbf{5.808} & \gap\gap 0.0000 \\
aBARCH-NN(15) &\gap 0.508 &\gap\gap 0.6119 \\
aBARCH-NN(35) &\gap 1.399 &\gap\gap 0.1629 \\
aBARCH-NN(55) &\gap {-2.607} & \gap\gap 0.0097 \\
aBARCH-NN(75) &\gap -0.812 &\gap\gap 0.4174 \\
aBARCH-NN(CO) &\gap \textbf{5.706} & \gap\gap 0.0000 \\
aSVR-GARCH &\gap 0.431 &\gap\gap 0.6666 \\
		\hline
	\end{tabular}
	\caption{\textbf{S\&P500}: DM test on S\&P500 data (benchmark is the SVR-GARCH result).}
	\label{fig:sp500_out-of-sample-mdtest}
\end{table}
\subsection{SH300}

We first depict typical behaviors on the test data set of SH300 \footnote{Only 100 out of 252 trading days are shown in this and the following figures to save space.} for different models as shown in Figure~\ref{fig:typical-behavior} where a GARCH model predicts well for the trending of the volatility; however, the performance is low since it cannot reveal the details of the volatility trending as shown in Figure~\ref{fig:typical_garch}. Figure~\ref{fig:typical_svrgarch} shows a typical graphical representation of the SVR-GARCH model where we may think it predicts well at first glance. However, 
the disadvantage of ``backward eavesdropping" can be easily observed, i.e., predict volatility at time $t$ by deviating the realized volatility at time $t-1$ to some extent. And the error is accumulated during the time period. 
The BARCH result shown in Figure~\ref{fig:typical_barch} partly solves the ``backward eavesdropping" problem and the RMSE of the BARCH-NN model is the smallest among the four. 
While, similar to the GARCH model, BARCH can also lose some detailed information, e.g., the prediction between 80-th and 100-th day in Figure~\ref{fig:typical_barch}. This problem is less sever in the BARCH-NN model which is because the neural network is a more powerful machine learning tool that can approximate universal functions \citep{gelenbe1999function}.

Figure~\ref{fig:sh300-typical-er} compares the augmented version of each model on SH300 data. In all comparisons, we find the augmented method can reduce underestimation during the peak areas (orange cycles in the figures) and reduce overestimation during the trough areas (cyan cycles in the figures) to some extent. 
As discussed in the introduction section, in real quantitative strategies, trading opportunities always happen in these areas. For example, in option trading, traders tend to sell option if the implies volatility is higher than the predicted volatility (i.e., the proxy of realized volatility (RV)) during the peak area. If the predicted volatility is underestimated, the strategy can lose money.

Though the BARCH-NN(55) and aBARCH-NN(55) shown in Figure~\ref{fig:sh300-typical-er2} are not the best ones we have obtained for the BARCH-NN models (we shall come back to the best one in Figure~\ref{fig:corr-sh300} in next paragraph), a comparison with the SVR-GARCH result shows the BARCH-NN performs better when the volatility is in the rising edge or descending edge (red arrows in Figure~\ref{fig:sh300-typical-er2} and Figure~\ref{fig:sh300-typical-er3}). 

As discussed previously, a correlation-based procedure is used to select the features for BARCH and BARCH-NN models where 10 out of 90 features are selected. The 
BARCH-NN(CO) and aBARCH-NN(CO) perform best for the SH300 data set. Figure~\ref{fig:corr-sh300} shows the difference between the predicted volatility and the realized volatility is small, and the ``backward eavesdropping" problem is solved as well. 

Table~\ref{fig:sh300_out-of-sample} reports the detailed forecasting volatility performance of different models for SH300 data in terms of RMSE and MAE based on Eq.~\eqref{equation:rmse} and \eqref{equation:mae}. We observe the BARCH-NN(CO) is close to the realized volatility, the RMSE and MAE performances also show promising results for the models. 
Since we mentioned the SVR-GARCH has the ``backward eavesdropping" problem, we also report the RMSE and MAE by simply predicting the volatility at time $t$ by that at time $t-1$. The measure is termed \textit{Eavesdrop} in Table~\ref{fig:sh300_out-of-sample}. We notice the SVR-GARCH is only a litter better than the \textit{Eavesdrop} result in terms of RMSE, while even worse than the latter in the sense of MAE. However, the proposed BARCH-NN(CO) or aBARCH-NN(CO) models are better in both of the two measurements.
Table~\ref{fig:sh300_out-of-sample-mdtest} provides the DM test of the BARCH models (benchmark is the SVR-GARCH result) where the feature selected BARCH-NN(CO) and aBARCH-NN(CO) models indicate significant differences to the SVR-GARCH model.

\subsection{S\&P500}
Similarly, Figure~\ref{fig:sp500-typical-er} compares the augmented version of each model on S\&P500. Again, in all comparisons, we find the augmented method can reduce underestimation during the peak areas (orange cycles in the figures) and reduce overestimation during the trough areas (cyan cycles in the figures) to some extent. Though the BARCH-NN(15) and aBARCH-NN(15) shown in Figure~\ref{fig:sp500-typical-e2} are not the best ones we have obtained for the BARCH-NN models, a comparison with the SVR-GARCH shows the BARCH-NN performs better when the volatility is in the rising edge or descending edge (red arrows in Figure~\ref{fig:sp500-typical-e2} and Figure~\ref{fig:sp500-typical-er3}). 

Same as the SH300 case, a correlation-based procedure is used to select the features for BARCH and BARCH-NN models where the same 10 out of 90 features are selected. This is partly because the ARCH, GARCH, EGARCH, or GJR models are consistent over different data sets. 
The BARCH-NN(CO) and aBARCH-NN(CO) perform best for the S\&P500 data set. Figure~\ref{fig:corr-sp500} shows the difference between the predicted volatility and the realized volatility is small for our best model, and again the ``backward eavesdropping" problem is also not observed on the S\&P500 data. 

Table~\ref{fig:sp500_out-of-sample} reports the detailed forecasting volatility performance of different models for S\&P500 data in terms of RMSE and MAE. 
Again, we observe that the SVR-GARCH model only performs a litter better than the \textit{Eavesdrop} result. While the proposed BARCH-NN(CO) or aBARCH-NN(CO) models are much better.
Table~\ref{fig:sp500_out-of-sample-mdtest} provides the DM test of the BARCH models (benchmark is the SVR-GARCH result). The feature selected BARCH-NN(CO) and aBARCH-NN(CO) models on S\&P500 data still provide evidence of significant difference to the SVR-GARCH model.

\section{Conclusion}
The aim of this paper is to solve the ``backward eavesdropping" problem in the SVR-GARCH model and overcome the overestimation and underestimation in various volatility prediction models. To check the mentioned problems and the performance of the proposed models, real market data sets including SH300 and S\&P500 are used in empirical analysis.
The article proposes a blending-ARCH model based on correlation feature selection that improves the performance in terms of RMSE, MAE, DM test, and pictorial behavior. 
Furthermore, the existing ARCH family or SVR-GARCH models tend to overestimate or underestimate in the peak and trough areas of the volatility sequence. The overestimation and underestimation problem is server in real quantitative trades since many trading strategies tend to bid or ask in the peak and trough areas of the volatility sequence. An augmented method based on the effective ratio can be applied to any existing volatility models and shows it can reduce overestimating and underestimating. 

\small
\bibliography{bibliography}
\bibliographystyle{sty}

\appendix

\section{Neural network structure}\label{appendix:-mpstructure}
For the network structure, we only use a structure with fully connected layers. For each fully connected layer, we denote it by F$(< \text{num outputs} >:< \text{activation function} >)$. Then the network structure we use can be described by:
\begin{equation}
\text{F(100:Relu) - F(50:Relu) - F(50:Relu) - F(1:MSE)}.
\end{equation}
And the parameter to optimize the network is given as follows:
\begin{itemize}
\item L2 penalty (regularization term): $\alpha$=0.0001;
\item Learning rate: 0.001;
\item Batch size: 200;
\item Optimizer: Adam($\beta_1=0.9, \beta_2=0.999$) \citep{kingma2014adam};

\end{itemize}

\section{Goodness of fit for SH300 and S\&P500}\label{appendix:goodness-fit}
In this section, we report the goodness of fit for SH300 and S\&P500 data sets under ARCH, GARCH, EGARCH, and GJR respectively with different innovations (the normal, the Student's t, the skewed Student's t, and the generalized error distribution).

\begin{table}[!h]
	\small
	\begin{tabular}{llll}
		\hline
		Model        &  LL         & AIC &  BIC\\ 
		\hline
		ARCH-N(14)      & 4937.811 & -9.906 & -9.995\\
		ARCH-t(11)     &   4837.193 & -9.700 & -9.778  \\ 
		ARCH-st(11)     & 4836.465 & -9.701 & -9.785    \\ 
		ARCH-G(10)     & 4835.353 & -9.695 & -9.767\\ 
		GARCH-N(1,1)     &  4925.859 & -9.858 & -9.876  \\ 
		GARCH-t(1,1)     &   4824.549 & -9.657 & -9.681 \\ 
		GARCH-st(1,1)     &  4823.814 & -9.658 & -9.688   \\ 
		GARCH-G(1,1)    &  4817.683 & -9.643 & -9.667 \\ 
		EGARCH-N(1,1)      & 4925.239 & -9.856 & -9.874  \\ 
		EGARCH-t(1,1)      & 4821.969 & -9.652 & -9.676   \\ 
		EGARCH-st(1,1)     & 4821.278 & -9.653 & -9.683\\ 
		EGARCH-G(1,1)     &  4815.950 & -9.640 & -9.664 \\ 
		GJR-N(1,1)     &   4925.644 & -9.859 & -9.883  \\ 
		GJR-t(1,1)     &   4819.234 & -9.650 & -9.686 \\ 
		GJR-st(2,1)     & 4818.685 & -9.651 & -9.693  \\ 
		GJR-G(1,1)     &4817.034 & -9.644 & -9.674 \\ 
		\hline
	\end{tabular}
	\caption{Out-of-sample goodness of fit on SH300 data. LL is the log likelihood, AIC is the Akaike information criterion, and BIC is the Bayesian information criterion.}
	\label{fig:sh300_goodness}
\end{table}

\begin{table}[!h]
		\small
	\begin{tabular}{llll}
		\hline
		Model        &  LL         & AIC &  BIC\\ 
		\hline
		
		ARCH-N(6)      &  3395.889 & -6.806 & -6.847   \\
		ARCH-t(6)     &  3350.526 & -6.717 & -6.764\\ 
		ARCH-st(6)     & 3349.848 & -6.718 & -6.771  \\ 
		ARCH-G(6)     &  3337.130 & -6.690 & -6.738\\ 
		GARCH-N(2,1)     &  3395.348 & -6.799 & -6.822\\ 
		GARCH-t(2,1)     &  3349.800 & -6.710 & -6.739 \\ 
		GARCH-st(2,1)     &  3349.082 & -6.710 & -6.746  \\ 
		GARCH-G(2,1)     &  3336.435 & -6.683 & -6.713 \\ 
		EGARCH-N(2,1)     & 3398.249 & -6.804 & -6.828  \\ 
		EGARCH-t(2,1)     & 3351.507 & -6.713 & -6.743  \\ 
		EGARCH-st(2,1)     & 3350.735 & -6.713 & -6.749   \\ 
		EGARCH-G(2,1)     &  3337.884 & -6.686 & -6.715  \\ 
		GJR-N(1,1)     &   3331.354 & -6.671 & -6.694\\ 
		GJR-t(1,1)     &  3299.961 & -6.610 & -6.640\\ 
		GJR-st(1,1)     &  3299.785 & -6.612 & -6.647  \\ 
		GJR-G(1,1)     &  3289.422 & -6.589 & -6.618 \\ 
		\hline
	\end{tabular}
	\caption{Out-of-sample goodness of fit on S\&P500 data.}
	\label{fig:sp500_goodness}
\end{table}

\end{document}